\documentclass[%
 reprint,
superscriptaddress,
 aps,
 prl,
floatfix,
]{revtex4-2}

\usepackage{graphicx}% Include figure files
\usepackage{dcolumn}% Align table columns on decimal point
\usepackage{bm}% bold math
\usepackage{amsmath,amssymb}
\usepackage{hyperref}% add hypertext capabilities
\hypersetup{
	colorlinks=true,
	linkcolor=blue,
	filecolor=blue,
	urlcolor=blue,
	citecolor=blue,
}

\begin{document}

% \preprint{APS/123-QED}

\title{Metastable confinement in Rydberg lattice gauge theories}

\author{Yaohua Li}
% \altaffiliation[Current address: ]{Niels Bohr Institute, University of Copenhagen, Jagtvej 155A, DK-2200 Copenhagen, Denmark}
\affiliation{Niels Bohr Institute, University of Copenhagen, Jagtvej 155A, DK-2200 Copenhagen, Denmark}
\affiliation{State Key Laboratory of Low-Dimensional Quantum Physics, Department of Physics, Tsinghua University, Beijing 100084, P. R. China
}%
\author{Devendra Singh Bhakuni}%
\affiliation{The Abdus Salam International Centre for Theoretical Physics (ICTP), Strada Costiera 11, Trieste 34151, Italy
}%
\author{Yong-Chun Liu}
\affiliation{State Key Laboratory of Low-Dimensional Quantum Physics, Department of Physics, Tsinghua University, Beijing 100084, P. R. China
}%
\affiliation{Frontier Science Center for Quantum Information, Beijing 100084, China}
\author{Marcello Dalmonte}
\affiliation{%
 The Abdus Salam International Centre for Theoretical Physics (ICTP), Strada Costiera 11, Trieste 34151, Italy
}%
\affiliation{Dipartimento di Fisica e Astronomia, Università di Bologna, via Irnerio 46, I-40126 Bologna, Italy}

\begin{abstract}
Confinement and string breaking are two fundamental phenomena in gauge theories. Signatures of both are currently pursued in quantum-simulator experiments, opening a new angle on strongly interacting dynamics of gauge fields out of equilibrium, complementary to traditional particle-physics settings. In this work, we report the emergence of metastable confinement dynamics in a U(1) lattice gauge theory, originating from the competition between string tension and four-Fermi coupling - a competition that naturally arises in Rydberg atom arrays. We show that the initial string state can be resonantly melted through controlled energy matching, a phenomenon we identify as resonant string breaking. We demonstrate this mechanism for both static and Floquet-driven systems, where periodic modulation generates a spectrum of tunable sideband resonances. 
Our work provides new insights into the mechanisms of confinement and string breaking driven by long-range interactions and time-dependent fields, which are available in current quantum simulators on a variety of platforms.
\end{abstract}

\maketitle

{\it Introduction.}--- Lattice gauge theories (LGTs) lie at the heart of our understanding of strongly correlated quantum systems, ranging from quark confinement in high-energy physics to the emergence of topological order in condensed matter~\cite{kogut_hamiltonian_1975,kogut_introduction_1979, Raifeartaigh2000gauge,hermele2004pyrochlore,levin2005colloquium,sachdev_topological_2018}. Paradigmatic features of these theories are confinement and string breaking \cite{GLIOZZIA199976,Philipsen1998string,Bali2005observation,Alkofer_2007,hebenstreit_real-time_2013,kuhn_non-abelian_2015,pichler_real-time_2016,Kormos2017,verdel2020real,verdel_dynamical_2023}, that dictate the non-equilibrium dynamics of quarks and composite objects such as mesons and baryons. 
However, classical simulation of real-time dynamics is very challenging due to factors such as complex action and sign problems~\cite{troyer2005computational,wiese_ultracold_2013,zohar_quantum_2015,Carmen_2020}. In recent years, quantum simulation has emerged as a compelling alternative, enabled by the realization of effective Hamiltonian that respect local gauge symmetries~\cite{Bloch2012,banerjee_atomic_2012,zohar_simulating_2012,jordan_quantum_2012,tagliacozzo_optical_2013,Banuls2020,Klco_2022,aidelsburger2022cold,funcke2022towards,Halimeh2025,bauer2023quantum}, as seen in systems with trapped ions~\cite{martinez_real-time_2016,muschik2017u,kokail_self-verifying_2019,tan_domain-wall_2021,meth_simulating_2025,mueller_quantum_2025,surace2024string,de2024observation}, superconducting qubits \cite{PhysRevResearch.4.L022060,cobos2025realtimedynamics21dgauge,cochran_visualizing_2025,mildenberger_confinement_2025}, optical lattices \cite{Banerjee2012atomic,gorg_realization_2019,schweizer_floquet_2019,mil_scalable_2020,yang_observation_2020,frolian_realizing_2022,Surace2023abinitio,zhou_thermalization_2022,zhu2024probing,Zhang2025} and Rydberg atom arrays~\cite{Bernien2017,Shaw2024,surace_lattice_2020,mark2025observation,González-Cuadra2025,datla2025statistical,xiang2025realtimescattering,mark2025observation}. In particular, Rydberg atom arrays naturally realize a U(1) gauge symmetry in the Rydberg blockade regime, where strong nearest-neighbor interactions impose a local constraint equivalent to Gauss’s law~\cite{glaetzle2014quantum,surace_lattice_2020}, enabling the experimental observation of confinement and string breaking~\cite{Bernien2017,González-Cuadra2025,xiang2025realtimescattering,mark2025observation}. However, the precise relationship between confinement and string-breaking in such time-dependent dynamics remains poorly understood, especially at intermediate times (which, for gauge theories, have been shown to host unexpected dynamical phenomena~\cite{Bernien2017,turner2018weak,xiang2025realtimescattering,mark2025observation,Birnkammer2022}).

In this work, we identify a metastable confinement regime in a U(1) LGT realized in a one-dimensional Rydberg array with engineered next-nearest-neighbor (NNN) interactions and staggered detuning. In this regime, the confinement of the initial string state ultimately melts into the thermal equilibrium of a gas of quark-antiquark pairs. Our observations can also be construed in the context of prethermalization, due to the emergence of (approximate) conserved charges in the system~\cite{Chen2021Emergent}, which are further apparent once reformulated within a solely fermionic Hilbert space~\cite{bhakuni2025anomalously}.

We further show that resonant melting occurs when the string tension (staggered detuning) and four-Fermi coupling (NNN interactions) become resonant. This resonant-string-breaking induces ergodic dynamics within resonant subspaces, which appear as sharp peaks in local observables. Finally, we extend this mechanism to Floquet‑driven systems, where periodic modulation generates tunable sideband resonances, providing a practical route to control and probe resonant dynamics in quantum simulators.

{\it Rydberg atom arrays to lattice gauge theory.}---
We consider a one-dimensional array of Rydberg atoms, where each atom is in an optical trap, equivalent to a qubit, with the ground state and Rydberg state mapped to $|0\rangle$ and $|1\rangle$, respectively. The Hamiltonian reads ($\hbar=1$)
\begin{eqnarray}\label{eq:ha}
    \hat{H}=\frac{\Omega}{2}\sum_{j}\hat{\sigma}_{j}^{x} - \sum_{j} \left[\Delta +(-1)^{j}\delta_{0} \right]\hat{n}_{j} + \sum_{j<k}V_{j,k}\hat{n}_{j}\hat{n}_{k},\ \
\end{eqnarray}
where $\hat{\sigma}_{j}^{\alpha}$ are Pauli matrices at site $j$, $\hat{n}_{j}=(\hat{\sigma}_{j}^{z}+1)/2$ is Rydberg excitation number operator at site $j$, $\Omega$ is the Rabi frequency, $\Delta $ and $\delta_{0} $ are global and staggered detunings, and $V_{j,k}=C_{6}/r_{j,k}^{6}$ is the Rydberg interaction strength that decays rapidly as distance $r_{j,k}$ between two Rydberg atoms at site $(j, k)$ increases. The NNN interactions can be enhanced by employing a staggered geometry, such as a one-dimensional zigzag ladder, which reduces the NNN distance. In two dimensions, lattice arrangements naturally yield similar staggered structures, leading to enhanced NNN couplings (see end matter).

\begin{figure*}[t]
    \centering
    \includegraphics[width=0.98\textwidth]{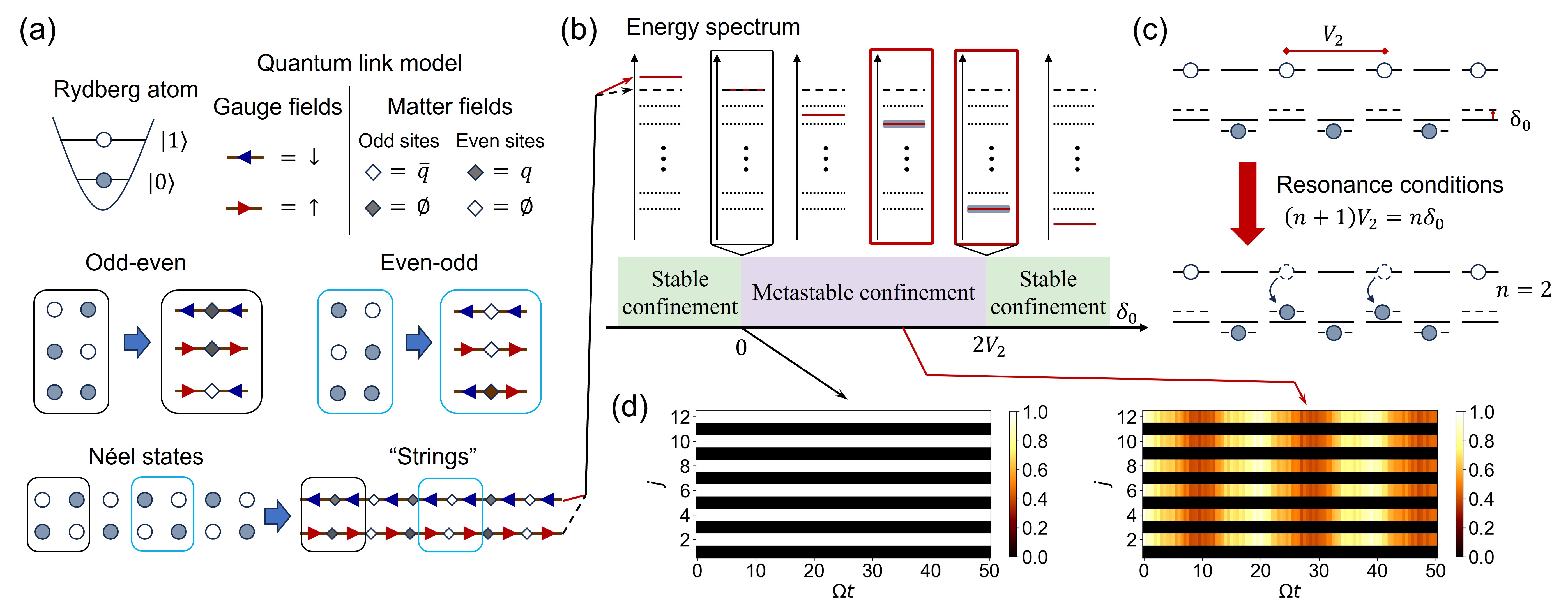}
    \caption{Schematics of LGTs description and resonant string breaking of Rydberg atom arrays. (a) Mapping between Rydberg atoms (with $ |0\rangle,|1\rangle$ being the ground and Rydberg state, respectively) and a U(1) quantum link model. The N\'{e}el states map into string states. (b) Schematic of stable and metastable confinement regimes. When the chosen initial state (N\'{e}el state shown in red solid line) lies at the extreme of the spectrum, the dynamics mimic the thermal equilibrium in the blockade Hilbert space, and both states show confinement. Conversely, when the initial state lies in the middle of the spectrum, the confinement will break in the thermal equilibrium.
    (c) Schematic of a resonant melting process starting from a N\'{e}el state where the resonance condition $(n+1)V_{2}=n\delta_{0}$ with $n=2$ is satisfied. (d) Time evolution of the average occupation $\langle n_{j}\rangle$ away from resonance ($\delta_{0}=0$, left) and at resonance ($\delta_{0}\approx 3V_{2}/2$, right).}
    \label{fig:skecth}
\end{figure*}

When the nearest interaction strength $V_{1}$ is much larger than all other energy scales ($\Omega$, $\Delta $, and $\delta_{0} $), two nearest atoms cannot be excited to the Rydberg states together due to the Rydberg blockade condition within a blockade radius $R_{\mathrm{b}}=(C_{6}/\Omega)^{1/6}$. The Rydberg blockade imposes the constraint $n_{j}n_{j+1}=0$, leading to an emergent U(1) gauge symmetry~\cite{surace_lattice_2020}. Consequently, the Hamiltonian Eq. (\ref{eq:ha}) can be used to simulate the U(1) LGTs in the Rydberg blockade regime. In conventional LGTs, the long-range Rydberg interaction is neglected. Here, we consider a case where the nearest optical traps are closer, and the NNN interactions become comparable to other energy scales--which, as we show below, corresponds to four-Fermi coupling. Further long-range interactions are much smaller and can still be neglected. Considering the mapping of Ref. \cite{surace_lattice_2020}, we can write the LGT Hamiltonian as
\begin{align}
        \hat{H}_{\mathrm{LGT}}=&-w\sum_{j}\left(\hat{\Phi}_{j}^{\dag}\hat{U}_{j,j+1}\hat{\Phi}_{j+1}+\mathrm{H.c.}\right) \nonumber\\
        &+ (m+m^{\prime})\sum_{j}\left(-1\right)^{j}\hat{\Lambda}_{j} \nonumber- 2m^{\prime}\sum_{j} \hat{\Lambda}_{j+1}\hat{\Lambda}_{j} \nonumber\\        &+J\sum_{j}\left(\hat{E}_{j,j+1}+\frac{\pi-\theta}{2\pi}\right)^{2}.
\end{align}
where $w=-\Omega/2$ is the coupling strength between the gauge and the matter field, $m=-\Delta /2$ is the fermion mass, and $m^{\prime}=V_{2}/2$ denotes the matter field energy (four-Fermi coupling) from the NNN interactions $V_{2}=V_{j,j+2}$, which corresponds to the interactions of two nearest-neighbor matter fields. $J$ is the gauge field energy, and $\theta$ is the topological angle, and they satisfy: $\delta_{0} =J(\pi-\theta)/\pi\equiv \chi/2$, where $\chi$ corresponds to the string tension in the classical limit. The operators $\hat{\Phi}_{j}$ describe staggered fermions, with  $\hat{\Lambda}_{j} = \hat{\Phi}_{j}^{\dag}\hat{\Phi}_{j}$ is the fermionic occupation, $\hat{U}_{j,j+1}$ are parallel transporters, and $\hat{E}_{j,j+1}$ denote the U(1) gauge field that mediates the interactions between the two nearest matter sites. As shown in Fig. \ref{fig:skecth}(a), the main idea of quantum simulation is mapping two nearest atoms into two nearest bonds in the QLM. Then the Rydberg blockade can be mapped to Gauss's law, and the N\'{e}el state $|Z_{2}\rangle=|\circ\bullet\circ\bullet\cdots\rangle$ can be mapped into continuous strings~\cite{surace_lattice_2020}.

\begin{figure*}[t]
    \centering
    \includegraphics[width=0.98\textwidth]{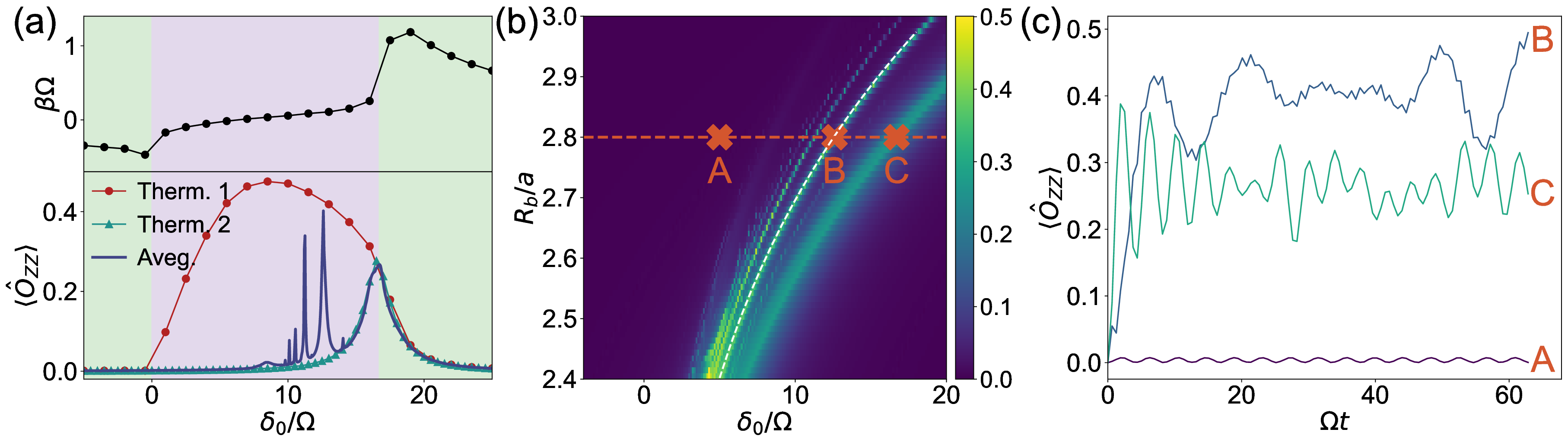}
    \caption{Thermal equilibrium and time average of local observables. (a) Upper panel: the effective temperature of the initial state. Lower panel: the thermal expectation values of the nearest-neighbor correlation $\hat{O}_{ZZ}$ in the blockade subspace (red, therm. 1) and in the resonant subspace (green, therm. 2), and the time-averaged expectation value of the observable (blue, aveg.). The blockade radius is $R_{\mathrm{b}}/a=2.8$ (corresponding to  $V_{2}/\Omega\approx8.11$). The light green (purple) area denotes the stable (metastable) confinement regime. (b) The nearest-neighbor correlation $\hat{O}_{ZZ}$ as a function of the staggered detuning $\delta_{0} $ and the blockade radius $R_{\mathrm{b}}$. The white dashed line corresponds to the resonant condition $3V_{2}\approx 2\delta_{0} $, and the red dashed line denotes the cut $R_{\mathrm{b}}/a=2.8$. (c) The time evolution of $\hat{O}_{ZZ}$ at the three characteristic points in (b). The number of atoms is $N=20$.}
    \label{fig:main}
\end{figure*}

{\it Metastable confinement and resonant melting.}---
The main finding of our work is illustrated in Fig. \ref{fig:skecth}(b). We are interested in the time evolution starting from a N\'{e}el (string) state, which has the largest energy from the four-Fermi coupling $V_{2}$. 
Here, we identify two distinct types of dynamics: stable and metastable confinement. The two regimes are distinguished by the relative position of the initial state, one of the N\'{e}el states, in the full spectrum (Fig. \ref{fig:skecth}(b)). When the initial state lies at the extreme of the spectrum (e.g., $\delta_0<0$), the occupations of other states in thermal equilibrium can be neglected, and confinement can be maintained in the long-time limit even when approaching thermal equilibrium. Oppositely, when the initial state lies inside the spectrum, the occupations of the other states in thermal equilibrium cannot be neglected. Before thermalization occurs, the dynamics remain slow and confining-like: we refer to this regime as metastable confinement.

Over metastable confinement, there are many resonances emerging between the matter-field energy $m$, $m^{\prime}$, and the electric-field energy $\chi$: each of those leads to string breaking, with the emergence of different numbers of particle-antiparticle pairs. For instance, the lowest-order resonance $2V_{2}=\Delta +\delta_{0} $ corresponds to the emergence of isolated defects ``$0100010$" from the N\'{e}el state, equivalent to a quark-antiquark pair on neighboring sites. Other resonances $(n+1)V_{2}=n(\Delta +\delta_{0}) $ with larger support correspond to multiple quark-antiquark pairs [see Fig. \ref{fig:skecth}(c) for $n=2$]. Below, we restrict to the case of zero mass $\Delta =0$; however, it is straightforward to generalize the results here to non-zero $\Delta$. We note that, for a non-zero mass $(\Delta \neq0)$ and zero four-fermi coupling ($V_2=0$), prethermal dynamics have been observed starting from a homogeneous gas of quark-antiquark pairs, due to the conserved meson number in confined systems~\cite{Birnkammer2022}.

In Fig. \ref{fig:skecth}(d), we depict the evolution of $n_i$ as a function of time:  the two regimes correspond to confinement dynamics away from resonance (left) and the string-breaking dynamics at resonances (right), both starting from an initial N\'{e}el state.

{\it Local observable.}---
To characterize the different confinement regimes, we consider the nearest-neighbor correlation described by the operator  $\hat{O}_{ZZ}=(1/N)\sum_{j}(1 - \hat{n}_{j})(1 - \hat{n}_{j+1})$
, which in the LGT corresponds to the average number of quarks and anti-quarks.

In the lower panel of Fig. \ref{fig:main}(a), we show the long-time [$1<\Omega t/(2\pi)<10$] average expectation value of the nearest-neighbor correlation (blue line, aveg.). We also compute the expectation value of the corresponding thermal ensemble in the blockade subspace (red line, therm. $1$ ), where the effective temperature is determined by the initial state (the upper part). These calculations are both carried out in the blockaded subspace, which is reasonable because the Rydberg blockade effect is extremely strong here (see end matter). 

We determine two distinct dynamical regimes. The light green regime denotes stable confinement, where the time-averaged expectation value closely matches the thermal expectation value. In this regime, the effective temperature $1/\beta$ is close to zero. It means that the initial state is either close to the ground state or the highest energy state, so the initial confinement is maintained in the thermal equilibrium. The purple regime denotes the metastable confinement regime, where the time-averaged expectation value is much smaller than the thermal expectation value. In this regime, the effective temperature $1/\beta$ is non-zero. We expect that the initial confinement will finally melt into a thermal equilibrium at a very long time ($t \sim e^{V_2/\Omega}$) with different values of observables. 

Several peaks in the time-averaged expectation values can be observed in the metastable confinement regime when the initial state becomes resonant with special types of excitations. In particular, we consider the widest peak at $2V_{2}=\delta_{0} $, which corresponds to the excitation islands ``0100010". The green line in Fig. \ref{fig:main}(a) (labeled as therm. $2$) represents the thermal expectation value in the resonant Hilbert space with the excitation islands ``0100010" (see end matter), and we find that it matches well with the time-averaged value. Furthermore, in Fig. \ref{fig:main}(b) we show the dependence of resonances on both the staggered detuning $\delta_{0} $ and the blockade radius $R_{\mathrm{b}}$. The white dashed line corresponds to another resonant condition $3V_{2}\approx 2\delta_{0}$, which also matches well with the peaks in numerical results. We also consider three different parameter points, one away from the resonance (A) and two at the resonance (B, C). As shown in Fig. \ref{fig:main}(c), the nearest-neighbor correlation remains small with a regular oscillation when the system is away from the resonance, and the nearest-neighbor correlations quickly obtain a large value and start irregular oscillations when the system is near resonances.

\begin{figure}[t]
    \centering
    \includegraphics[width=0.49\textwidth]{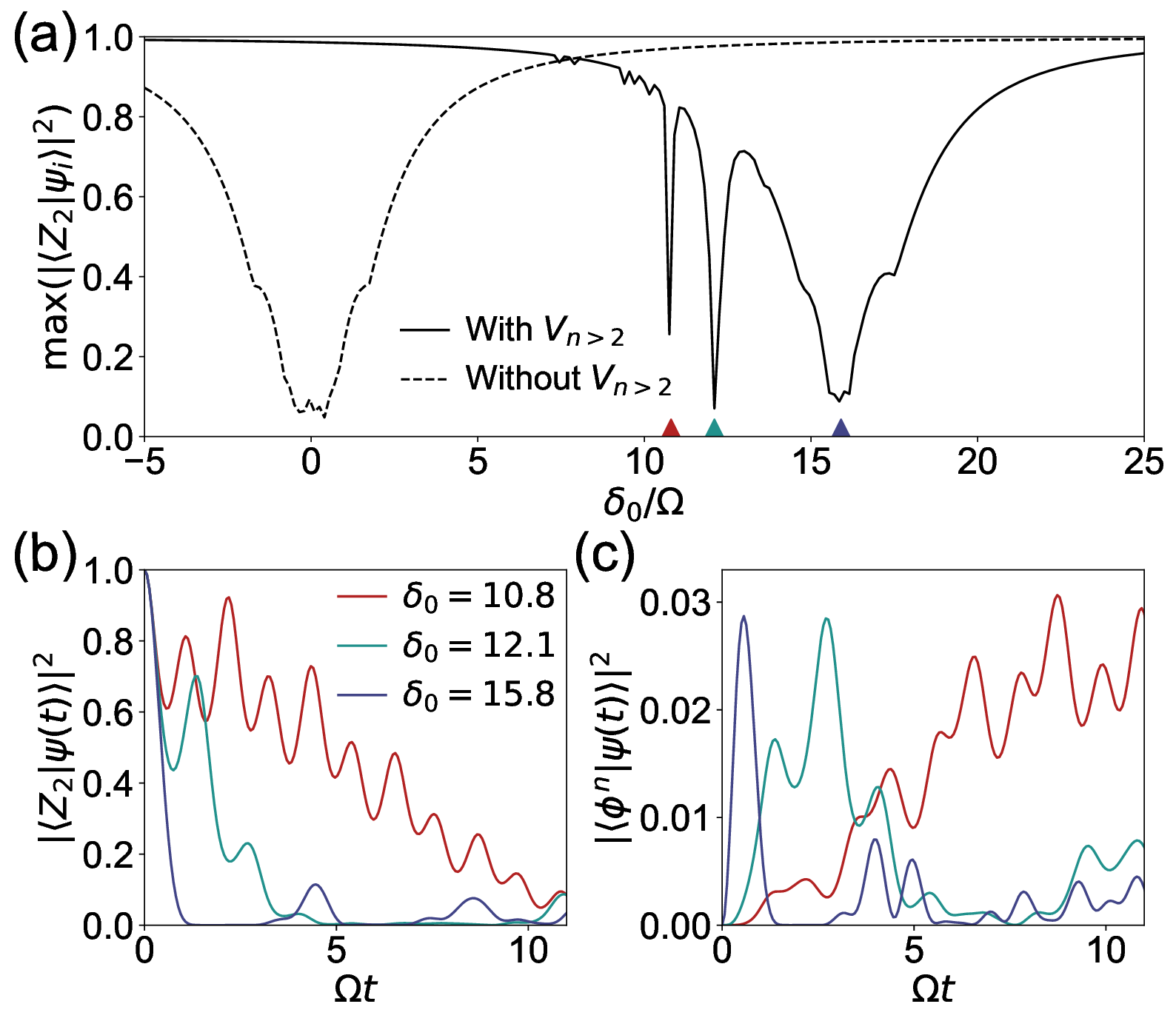}
    \caption{Resonant string breaking in the eigenstates and dynamics. (a) Maximal overlap between the eigenstates $|\psi_{i}\rangle$ and the initial state $|Z_{2}\rangle$. The three triangles denote the three resonances at $(n+1)V_{2}= n\delta_{0} $ for $n=1,2,3$, corresponding to the three lines of the same colors in (b) and (c). (b),(c) The overlaps between the dynamical state $|\psi(t)\rangle$ and the initial state $|Z_{2}\rangle$ in (b) or the one-island excitation state $|\phi^{n}\rangle$ in (c). $n$ denotes the number of quark-antiquark pairs in an isolated island.  The number of atoms is $N=28$.}
    \label{fig:overlap}
\end{figure}

{\it State overlaps.}---
Another evidence of resonances is the maximal overlap between the eigenstates $|\psi_{i}\rangle$ and the initial state $|Z_{2}\rangle$, as shown in Fig. \ref{fig:overlap}(a). When the staggered detuning is away from resonances, the maximal overlap is close to unity. It indicates a parametrically slow dynamics in both the stable and metastable confinement regimes, allowing the initial string state to be unaffected by the dynamics over a long timescale  \cite{yang_probing_2025}.

In contrast, the maximal overlap sharply decreases at the resonances. The initial string state $|Z_{2}\rangle$ is highly mixed with the other states near the resonances, leading to the resonant melting of the initial string state. The three resonances in Fig. \ref{fig:overlap}(a) represented by purple, blue, and red triangles correspond to the resonant conditions $(n+1)V_{2}= n\delta_{0} $ for $n=1,2,3$, respectively. They are related to the excitation of isolated islands with $n$ pairs of quarks-antiquarks per island. 
As a comparison, we also show the results of the PXP model without long-range interactions (dashed line), where all the resonances mix at $\delta_0=0$ and can not be distinguished.

\begin{figure}[b]
    \centering
    \includegraphics[width=0.49\textwidth]{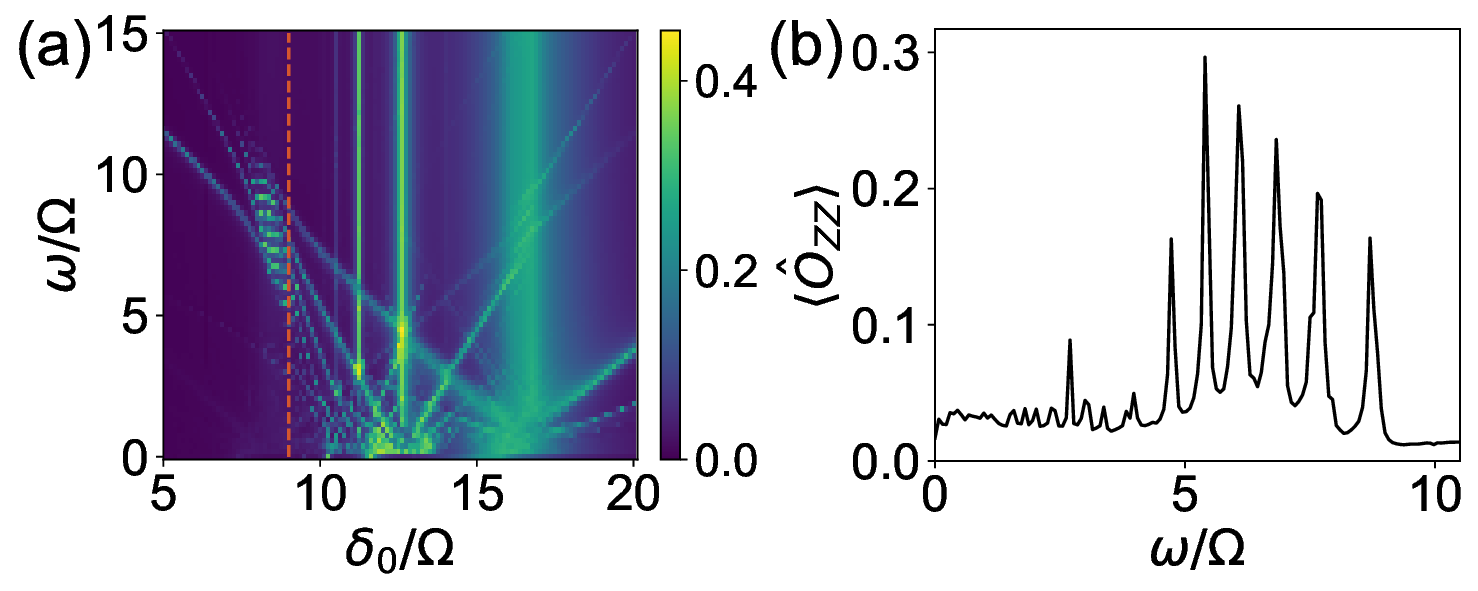}
    \caption{Floquet resonances from periodic modulation of global detuning $\Delta=\Delta_{\mathrm{m}}\cos(\omega t)$. (a) The nearest-neighbor correlation $\hat{O}_{ZZ}$ as a function of the staggered detuning $\delta_{0} $ and the modulation frequency $\omega$. (b) The detailed resonance structure for $\delta_{0}/\Omega=9$, corresponding to the red dashed line in (a).  The number of atoms is $N=20$.}
    \label{fig:floquet}
\end{figure}

We now discuss the time evolution of the overlaps for various values of string tension $\delta_{0}$ satisfying the resonance condition $\delta_0 = 10.8\Omega,\ 12.1\Omega,$ and $15.8\Omega$ starting from a Neel state. In Figs. \ref{fig:overlap}(b), we show the overlaps between the dynamical state $|\psi(t)\rangle$ and the initial state $|Z_{2}\rangle$; in Fig.~\ref{fig:overlap}(c), we instead show the overlap with the one-island excitation state that is resonant at that value of $\delta_0$ ($``0100010"$, $``010000010"$, and $``01000000010"$ for $n=1,2,3$, respectively).

The probabilities of the string state $|Z_{2}\rangle$ decrease linearly at the beginning, accompanied by a linear increase in the probability of the one-island excitation state $|\phi^{n}\rangle$. The rates of decrease or increase depend on the dimensions of the resonant Hilbert space. For the three resonances $n=1,2,3$, the dimensions are $123$, $43$, $26$ for a periodic Rydberg array with $28$ atoms. Larger dimensions lead to quicker resonant melting into the resonant subspace. For small subspace dimensions, such as the cases $n=2,3$, there are oscillations with partial revivals during the resonant melting process. These oscillations disappear for $n=1$ due to the short melting time. At longer times (not shown here), the probabilities of the one-island excitation state $|\phi^{n}\rangle$ also decrease with a similar rate, and the probabilities are further transferred to states with more isolated islands. 

We note that the maximum probability of the one-island excitation state $\langle\phi^{n}|\psi(t)\rangle^2$ is close to 0.03 because there are $N=28$ degenerate $|\phi^{n}\rangle$. Moreover, the dimensions of the resonant subspace also determine the resonant widths as shown in Fig. \ref{fig:main}(a) and Fig. \ref{fig:overlap}(a).

{\it Floquet resonances.}---We can also generalize the resonances between matter and electric fields to Floquet systems. In Fig. \ref{fig:floquet}(a), we study the nearest-neighbor correlation $\hat{O}_{ZZ}$ after introducing a periodic modulation of global detuning $\Delta=\Delta_{\mathrm{m}}\cos(\omega t)$. The main peaks satisfying $(n+1)V_{2}= n\delta_{0}$, remain unchanged. However, there are rich sidebands given by $\pm m\omega +(n+1)V_{2}= n\delta_{0} $ emerging from the main peaks on increasing the modulation frequency $\omega$. These sidebands demonstrate a method for controlling the resonances. We can separate these resonances or, inversely, make them close to each other. The former could be useful for experimental observation, as unavoidable fluctuations of Rydberg interaction strengths limit the resolution of two resonances.

We observe clear avoided crossing between two or more resonances, which is a signature of strong interactions and non-integrable dynamics in the Rydberg system. Fig. \ref{fig:floquet}(b) further shows the detailed structure of the Floquet resonances near the avoided crossing. The peaks are modulated by the couplings to each other and deviate from the resonant conditions $\pm m\omega +(n+1)V_{2}= n\delta_{0} $. Instead, they are nearly distributed at equal distances.

{\it Discussion.}---
We have demonstrated the emergence of a metastable confinement regime resulting from the competition between string tension and four-Fermi coupling in a U(1) lattice gauge theory. 
Such metastable confinement can be understood from the relative position of the initial string state in the full spectrum, and is characterized by an effective temperature.  When the effective temperature is away from zero, the metastable confinement will eventually melt into the thermal equilibrium of a gas of quark-antiquark pairs at a very long time.
Moreover, in the metastable regime, we observe resonant melting due to the resonant creation of quark-antiquark pairs at some special resonance conditions between the string tension and four-Fermi coupling.

This resonant string breaking manifests as peaks in local observables and drops in the overlap of the initial state with the eigenstates, signaling the onset of ergodic behavior within the resonant subspace.
We further extend it to Floquet-driven systems, showing that periodic modulation of the global detuning introduces a new set of sideband resonances with high tunability.

Our work highlights how engineered long-range interactions can be utilized to probe the competition between confinement and thermalization in current Rydberg atom experiments. It would be interesting to extend this treatment to non-Abelian gauge groups, which are promising candidates for richer resonance scenarios thanks to the presence of an extended meson and baryon spectrum.

We thank Prof. Klaus M{\o}lmer, Fan Yang, and Weilun Jiang for helpful discussions, and A. Bastianello for insightful comments on the manuscript. Y.L. and Y.-C.L. acknowledge support from the National Natural Science Foundation of China (NSFC) (Grant No. 123B2066, No. 92576204, No. 12275145). M.~D. was partly supported by the QUANTERA DYNAMITE PCI2022-132919, by the EU-Flagship programme Pasquans2, by the PNRR MUR project PE0000023-NQSTI, the PRIN programme (project CoQuS), and by the ERC Consolidator grant WaveNets (Grant agreement ID: 101087692). 

\bibliography{ref}% Produces the bibliography via BibTeX.

@Article{Birnkammer2022,
author={Birnkammer, Stefan
and Bastianello, Alvise
and Knap, Michael},
title={Prethermalization in one-dimensional quantum many-body systems with confinement},
journal={Nature Communications},
year={2022},
month={Dec},
day={10},
volume={13},
number={1},
pages={7663},
issn={2041-1723},
doi={10.1038/s41467-022-35301-6}
}

@article{troyer2005computational,
  title = {Computational Complexity and Fundamental Limitations to Fermionic Quantum Monte Carlo Simulations},
  author = {Troyer, Matthias and Wiese, Uwe-Jens},
  journal = {Phys. Rev. Lett.},
  volume = {94},
  issue = {17},
  pages = {170201},
  numpages = {4},
  year = {2005},
  month = {May},
  publisher = {American Physical Society},
  doi = {10.1103/PhysRevLett.94.170201}
}

@article{Carmen_2020,
doi = {10.1088/1361-6633/ab6311},
year = {2020},
month = {jan},
publisher = {IOP Publishing},
volume = {83},
number = {2},
pages = {024401},
author = {Carmen Bañuls, Mari and Cichy, Krzysztof},
title = {Review on novel methods for lattice gauge theories},
journal = {Reports on Progress in Physics},
}

@article{celi_emerging_2020,
	title = {Emerging {Two}-{Dimensional} {Gauge} {Theories} in {Rydberg} {Configurable} {Arrays}},
	volume = {10},
	issn = {2160-3308},
	url = {https://link.aps.org/doi/10.1103/PhysRevX.10.021057},
	number = {2},
	urldate = {2026-01-13},
	journal = {Phys. Rev. X},
	author = {Celi, Alessio and Vermersch, Benoît and Viyuela, Oscar and Pichler, Hannes and Lukin, Mikhail D. and Zoller, Peter},
	month = jun,
	year = {2020},
	pages = {021057},
}

@article{glaetzle2014quantum,
  title={Quantum spin-ice and dimer models with Rydberg atoms},
  author={Glaetzle, Alexander W and Dalmonte, Marcello and Nath, Rejish and Rousochatzakis, Ioannis and Moessner, Roderich and Zoller, Peter},
  journal={Physical Review X},
  volume={4},
  number={4},
  pages={041037},
  year={2014},
  publisher={APS}
}

@article{turner2018weak,
  title={Weak ergodicity breaking from quantum many-body scars},
  author={Turner, Christopher J and Michailidis, Alexios A and Abanin, Dmitry A and Serbyn, Maksym and Papi{\'c}, Zlatko},
  journal={Nature Physics},
  volume={14},
  number={7},
  pages={745--749},
  year={2018},
  publisher={Nature Publishing Group UK London}
}

@article{abanin_rigorous_2017,
	title = {A {Rigorous} {Theory} of {Many}-{Body} {Prethermalization} for {Periodically} {Driven} and {Closed} {Quantum} {Systems}},
	volume = {354},
	issn = {1432-0916},
	url = {https://doi.org/10.1007/s00220-017-2930-x},
	number = {3},
	urldate = {2026-01-12},
	journal = {Commun. Math. Phys.},
	author = {Abanin, Dmitry and De Roeck, Wojciech and Ho, Wen Wei and Huveneers, François},
	month = sep,
	year = {2017},
	pages = {809--827},
}

@article{mori_thermalization_2018,
	title = {Thermalization and prethermalization in isolated quantum systems: a theoretical overview},
	volume = {51},
	issn = {0953-4075},
	shorttitle = {Thermalization and prethermalization in isolated quantum systems},
	url = {https://doi.org/10.1088/1361-6455/aabcdf},
	number = {11},
	urldate = {2026-01-12},
	journal = {J. Phys. B: At. Mol. Opt. Phys.},
	author = {Mori, Takashi and Ikeda, Tatsuhiko N and Kaminishi, Eriko and Ueda, Masahito},
	month = may,
	year = {2018},
	pages = {112001},
}

@article{berges_prethermalization_2004,
	title = {Prethermalization},
	volume = {93},
	url = {https://link.aps.org/doi/10.1103/PhysRevLett.93.142002},
	number = {14},
	urldate = {2026-01-12},
	journal = {Phys. Rev. Lett.},
	author = {Berges, J. and Borsányi, Sz. and Wetterich, C.},
	month = sep,
	year = {2004},
	pages = {142002},
}

@article{wiese_ultracold_2013,
	title = {Ultracold quantum gases and lattice systems: quantum simulation of lattice gauge theories},
	volume = {525},
	copyright = {© 2013 by WILEY-VCH Verlag GmbH \& Co. KGaA, Weinheim},
	issn = {1521-3889},
	shorttitle = {Ultracold quantum gases and lattice systems},
	url = {https://onlinelibrary.wiley.com/doi/abs/10.1002/andp.201300104},
	number = {10-11},
	urldate = {2026-01-12},
	journal = {Ann. Phys. (Amsterdam)},
	author = {Wiese, U.-J.},
	year = {2013},
	pages = {777--796},
}

@article{hebenstreit_real-time_2013,
	title = {Real-{Time} {Dynamics} of {String} {Breaking}},
	volume = {111},
	url = {https://link.aps.org/doi/10.1103/PhysRevLett.111.201601},
	number = {20},
	urldate = {2026-01-12},
	journal = {Phys. Rev. Lett.},
	author = {Hebenstreit, F. and Berges, J. and Gelfand, D.},
	month = nov,
	year = {2013},
	pages = {201601},
}

@article{kuhn_non-abelian_2015,
	title = {Non-{Abelian} string breaking phenomena with matrix product states},
	volume = {2015},
	issn = {1029-8479},
	url = {https://doi.org/10.1007/JHEP07(2015)130},
	number = {7},
	urldate = {2026-01-12},
	journal = {J. High Energ. Phys.},
	author = {Kühn, Stefan and Zohar, Erez and Cirac, J. Ignacio and Bañuls, Mari Carmen},
	month = jul,
	year = {2015},
	pages = {130},
}

@article{pichler_real-time_2016,
	title = {Real-{Time} {Dynamics} in {U}(1) {Lattice} {Gauge} {Theories} with {Tensor} {Networks}},
	volume = {6},
	url = {https://link.aps.org/doi/10.1103/PhysRevX.6.011023},
	number = {1},
	urldate = {2026-01-12},
	journal = {Phys. Rev. X},
	author = {Pichler, T. and Dalmonte, M. and Rico, E. and Zoller, P. and Montangero, S.},
	month = mar,
	year = {2016},
	pages = {011023},
}

@article{verdel_dynamical_2023,
	title = {Dynamical {Localization} {Transition} of {String} {Breaking} in {Quantum} {Spin} {Chains}},
	volume = {131},
	url = {https://link.aps.org/doi/10.1103/PhysRevLett.131.230402},
	number = {23},
	urldate = {2026-01-12},
	journal = {Phys. Rev. Lett.},
	author = {Verdel, Roberto and Zhu, Guo-Yi and Heyl, Markus},
	month = dec,
	year = {2023},
	pages = {230402},
}

@misc{mark2025observation,
      title={Observation of ballistic plasma and memory in high-energy gauge theory dynamics}, 
      author={Daniel K. Mark and Federica M. Surace and Thomas Schuster and Adam L. Shaw and Wenjie Gong and Soonwon Choi and Manuel Endres},
      eprint={2510.11679}
}

@misc{bhakuni2025anomalously,
      title={Anomalously fast transport in non-integrable lattice gauge theories}, 
      author={Devendra Singh Bhakuni and Roberto Verdel and Jean-Yves Desaules and Maksym Serbyn and Marko Ljubotina and Marcello Dalmonte},
      year={2025},
      eprint={2509.08889},
      archivePrefix={arXiv}
}

@misc{datla2025statistical,
      title={Statistical Localization in a Rydberg Simulator of $U(1)$ Lattice Gauge Theory}, 
      author={Prithvi Raj Datla and Luheng Zhao and Wen Wei Ho and Natalie Klco and Huanqian Loh},
      eprint={2505.18143},
      archivePrefix={arXiv}
}

@Article{González-Cuadra2025,
author={Gonz{\'a}lez-Cuadra, Daniel
and Hamdan, Majd
and Zache, Torsten V.
and Braverman, Boris
and Kornja{\v{c}}a, Milan
and Lukin, Alexander
and Cant{\'u}, Sergio H.
and Liu, Fangli
and Wang, Sheng-Tao
and Keesling, Alexander
and Lukin, Mikhail D.
and Zoller, Peter
and Bylinskii, Alexei},
title={Observation of string breaking on a (2 + 1)D Rydberg quantum simulator},
journal={Nature},
year={2025},
month={Jun},
day={01},
volume={642},
number={8067},
pages={321-326},
issn={1476-4687},
doi={10.1038/s41586-025-09051-6}
}

@misc{xiang2025realtimescattering,
      title={Real-time scattering and freeze-out dynamics in Rydberg-atom lattice gauge theory}, 
      author={De-Sheng Xiang and Peng Zhou and Chang Liu and Hao-Xiang Liu and Yao-Wen Zhang and Dong Yuan and Kuan Zhang and Biao Xu and Marcello Dalmonte and Dong-Ling Deng and Lin Li},
      eprint={2508.06639},
      archivePrefix={arXiv}
}

@Article{Bernien2017,
author={Bernien, Hannes
and Schwartz, Sylvain
and Keesling, Alexander
and Levine, Harry
and Omran, Ahmed
and Pichler, Hannes
and Choi, Soonwon
and Zibrov, Alexander S.
and Endres, Manuel
and Greiner, Markus
and Vuleti{\'{c}}, Vladan
and Lukin, Mikhail D.},
title={Probing many-body dynamics on a 51-atom quantum simulator},
journal={Nature},
year={2017},
month={Nov},
day={01},
volume={551},
number={7682},
pages={579-584},
issn={1476-4687},
doi={10.1038/nature24622}
}

@Article{Shaw2024,
author={Shaw, Adam L.
and Chen, Zhuo
and Choi, Joonhee
and Mark, Daniel K.
and Scholl, Pascal
and Finkelstein, Ran
and Elben, Andreas
and Choi, Soonwon
and Endres, Manuel},
title={Benchmarking highly entangled states on a 60-atom analogue quantum simulator},
journal={Nature},
year={2024},
month={Apr},
day={01},
volume={628},
number={8006},
pages={71-77},
issn={1476-4687},
doi={10.1038/s41586-024-07173-x}
}

@article{Raifeartaigh2000gauge,
  title = {Gauge theory: Historical origins and some modern developments},
  author = {O'Raifeartaigh, Lochlainn and Straumann, Norbert},
  journal = {Rev. Mod. Phys.},
  volume = {72},
  issue = {1},
  pages = {1--23},
  numpages = {0},
  year = {2000},
  month = {Jan},
  publisher = {American Physical Society},
  doi = {10.1103/RevModPhys.72.1},
  url = {https://link.aps.org/doi/10.1103/RevModPhys.72.1}
}

@Article{Zhang2025,
author={Zhang, Wei-Yong
and Liu, Ying
and Cheng, Yanting
and He, Ming-Gen
and Wang, Han-Yi
and Wang, Tian-Yi
and Zhu, Zi-Hang
and Su, Guo-Xian
and Zhou, Zhao-Yu
and Zheng, Yong-Guang
and Sun, Hui
and Yang, Bing
and Hauke, Philipp
and Zheng, Wei
and Halimeh, Jad C.
and Yuan, Zhen-Sheng
and Pan, Jian-Wei},
title={Observation of microscopic confinement dynamics by a tunable topological $\theta$-angle},
journal={Nature Physics},
year={2025},
month={Jan},
day={01},
volume={21},
number={1},
pages={155-160},
issn={1745-2481},
doi={10.1038/s41567-024-02702-x}
}

@article{zhu2024probing,
  title={Probing false vacuum decay on a cold-atom gauge-theory quantum simulator},
  author={Zhu, Zi-Hang and Liu, Ying and Lagnese, Gianluca and Surace, Federica Maria and Zhang, Wei-Yong and He, Ming-Gen and Halimeh, Jad C and Dalmonte, Marcello and Morampudi, Siddhardh C and Wilczek, Frank and others},
  journal={arXiv:2411.12565},
  year={2024}
}

@article{de2024observation,
  title={Observation of string-breaking dynamics in a quantum simulator},
  author={De, Arinjoy and Lerose, Alessio and Luo, De and Surace, Federica M and Schuckert, Alexander and Bennewitz, Elizabeth R and Ware, Brayden and Morong, William and Collins, Kate S and Davoudi, Zohreh and others},
  journal={arXiv:2410.13815},
  year={2024}
}

@article{surace2024string,
  title={String-breaking dynamics in quantum adiabatic and diabatic processes},
  author={Surace, Federica Maria and Lerose, Alessio and Katz, Or and Bennewitz, Elizabeth R and Schuckert, Alexander and Luo, De and De, Arinjoy and Ware, Brayden and Morong, William and Collins, Kate and others},
  journal={arXiv:2411.10652},
  year={2024}
}

@Article{Halimeh2025,
author={Halimeh, Jad C.
and Aidelsburger, Monika
and Grusdt, Fabian
and Hauke, Philipp
and Yang, Bing},
title={Cold-atom quantum simulators of gauge theories},
journal={Nature Physics},
year={2025},
month={Jan},
day={01},
volume={21},
number={1},
pages={25-36},
issn={1745-2481},
doi={10.1038/s41567-024-02721-8}
}

@article{schweizer_floquet_2019,
	title = {Floquet approach to $\mathbb{Z}_{2}$ lattice gauge theories with ultracold atoms in optical lattices},
	volume = {15},
	issn = {1745-2481},
	url = {https://doi.org/10.1038/s41567-019-0649-7},
	doi = {10.1038/s41567-019-0649-7},
	number = {11},
	journal = {Nat. Phys.},
	author = {Schweizer, Christian and Grusdt, Fabian and Berngruber, Moritz and Barbiero, Luca and Demler, Eugene and Goldman, Nathan and Bloch, Immanuel and Aidelsburger, Monika},
	month = nov,
	year = {2019},
	pages = {1168--1173},
}

@article{gorg_realization_2019,
	title = {Realization of density-dependent {Peierls} phases to engineer quantized gauge fields coupled to ultracold matter},
	volume = {15},
	issn = {1745-2481},
	url = {https://doi.org/10.1038/s41567-019-0615-4},
	doi = {10.1038/s41567-019-0615-4},
	number = {11},
	journal = {Nat. Phys.},
	author = {G{\" o}rg, Frederik and Sandholzer, Kilian and Minguzzi, Joaquín and Desbuquois, Rémi and Messer, Michael and Esslinger, Tilman},
	month = nov,
	year = {2019},
	pages = {1161--1167},
}

@article{yang_observation_2020,
	title = {Observation of gauge invariance in a 71-site {Bose}–{Hubbard} quantum simulator},
	volume = {587},
	issn = {1476-4687},
	url = {https://doi.org/10.1038/s41586-020-2910-8},
	doi = {10.1038/s41586-020-2910-8},
	number = {7834},
	journal = {Nature},
	author = {Yang, Bing and Sun, Hui and Ott, Robert and Wang, Han-Yi and Zache, Torsten V. and Halimeh, Jad C. and Yuan, Zhen-Sheng and Hauke, Philipp and Pan, Jian-Wei},
	month = nov,
	year = {2020},
	pages = {392--396},
}

@article{mil_scalable_2020,
	title = {A scalable realization of local {U}(1) gauge invariance in cold atomic mixtures},
	volume = {367},
	url = {https://www.science.org/doi/abs/10.1126/science.aaz5312},
	doi = {10.1126/science.aaz5312},
	number = {6482},
	journal = {Science},
	author = {Mil, Alexander and Zache, Torsten V. and Hegde, Apoorva and Xia, Andy and Bhatt, Rohit P. and Oberthaler, Markus K. and Hauke, Philipp and Berges, Jürgen and Jendrzejewski, Fred},
	year = {2020},
	pages = {1128--1130},
}

@article{frolian_realizing_2022,
	title = {Realizing a {1D} topological gauge theory in an optically dressed {BEC}},
	volume = {608},
	issn = {1476-4687},
	url = {https://doi.org/10.1038/s41586-022-04943-3},
	doi = {10.1038/s41586-022-04943-3},
	number = {7922},
	journal = {Nature},
	author = {Fr{\" o}lian, Anika and Chisholm, Craig S. and Neri, Elettra and Cabrera, Cesar R. and Ramos, Ramón and Celi, Alessio and Tarruell, Leticia},
	month = aug,
	year = {2022},
	pages = {293--297},
}

@article{zhou_thermalization_2022,
	title = {Thermalization dynamics of a gauge theory on a quantum simulator},
	volume = {377},
	url = {https://www.science.org/doi/abs/10.1126/science.abl6277},
	doi = {10.1126/science.abl6277},
	number = {6603},
	journal = {Science},
	author = {Zhou, Zhao-Yu and Su, Guo-Xian and Halimeh, Jad C. and Ott, Robert and Sun, Hui and Hauke, Philipp and Yang, Bing and Yuan, Zhen-Sheng and Berges, Jürgen and Pan, Jian-Wei},
	year = {2022},
	pages = {311--314},
}

@article{mildenberger_confinement_2025,
	title = {Confinement in a $\mathbb{Z}_{2}$ lattice gauge theory on a quantum computer},
	volume = {21},
	issn = {1745-2481},
	url = {https://doi.org/10.1038/s41567-024-02723-6},
	doi = {10.1038/s41567-024-02723-6},
	number = {2},
	journal = {Nat. Phys.},
	author = {Mildenberger, Julius and Mruczkiewicz, Wojciech and Halimeh, Jad C. and Jiang, Zhang and Hauke, Philipp},
	month = feb,
	year = {2025},
	pages = {312--317},
}

@article{cochran_visualizing_2025,
	title = {Visualizing dynamics of charges and strings in (2 {+} 1){D} lattice gauge theories},
	volume = {642},
	issn = {1476-4687},
	url = {https://doi.org/10.1038/s41586-025-08999-9},
	doi = {10.1038/s41586-025-08999-9},
	number = {8067},
	journal = {Nature},
	author={Cochran, Tyler A and Jobst, Bernhard and Rosenberg, Eliott and Lensky, Yuri D and Gyawali, Gaurav and Eassa, Norhan and Will, Melissa and Szasz, Aaron and Abanin, Dmitry and Acharya, Rajeev and others},
	month = jun,
	year = {2025},
	pages = {315--320},
}

@article{muschik2017u,
  title={U (1) Wilson lattice gauge theories in digital quantum simulators},
  author={Muschik, Christine and Heyl, Markus and Martinez, Esteban and Monz, Thomas and Schindler, Philipp and Vogell, Berit and Dalmonte, Marcello and Hauke, Philipp and Blatt, Rainer and Zoller, Peter},
  journal={New Journal of Physics},
  volume={19},
  number={10},
  pages={103020},
  year={2017}
}

@article{levin2005colloquium,
  title = {Colloquium: Photons and electrons as emergent phenomena},
  author = {Levin, Michael and Wen, Xiao-Gang},
  journal = {Rev. Mod. Phys.},
  volume = {77},
  issue = {3},
  pages = {871--879},
  numpages = {0},
  year = {2005},
  month = {Sep},
  publisher = {American Physical Society},
  doi = {10.1103/RevModPhys.77.871}
}

@article{hermele2004pyrochlore,
  title = {Pyrochlore photons: The $U(1)$ spin liquid in a $S=\frac{1}{2}$ three-dimensional frustrated magnet},
  author = {Hermele, Michael and Fisher, Matthew P. A. and Balents, Leon},
  journal = {Phys. Rev. B},
  volume = {69},
  issue = {6},
  pages = {064404},
  numpages = {21},
  year = {2004},
  month = {Feb},
  publisher = {American Physical Society},
  doi = {10.1103/PhysRevB.69.064404}
}

@article{Banerjee2012atomic,
  title = {Atomic Quantum Simulation of Dynamical Gauge Fields Coupled to Fermionic Matter: From String Breaking to Evolution after a Quench},
  author = {Banerjee, D. and Dalmonte, M. and M\"uller, M. and Rico, E. and Stebler, P. and Wiese, U.-J. and Zoller, P.},
  journal = {Phys. Rev. Lett.},
  volume = {109},
  issue = {17},
  pages = {175302},
  numpages = {5},
  year = {2012},
  month = {Oct},
  publisher = {American Physical Society},
  doi = {10.1103/PhysRevLett.109.175302}
}

@article{Surace2023abinitio,
  title = {Ab Initio Derivation of Lattice-Gauge-Theory Dynamics for Cold Gases in Optical Lattices},
  author = {Surace, Federica Maria and Fromholz, Pierre and Oppong, Nelson Darkwah and Dalmonte, Marcello and Aidelsburger, Monika},
  journal = {PRX Quantum},
  volume = {4},
  issue = {2},
  pages = {020330},
  numpages = {18},
  year = {2023},
  month = {May},
  publisher = {American Physical Society},
  doi = {10.1103/PRXQuantum.4.020330}
}

@misc{cobos2025realtimedynamics21dgauge,
      title={Real-Time Dynamics in a (2+1)-D Gauge Theory: The Stringy Nature on a Superconducting Quantum Simulator}, 
      author={Jesús Cobos and Joana Fraxanet and César Benito and Francesco di Marcantonio and Pedro Rivero and Kornél Kapás and Miklós Antal Werner and {\" O}rs Legeza and Alejandro Bermudez and Enrique Rico},
      eprint={2507.08088},
      archivePrefix={arXiv},
}

@article{PhysRevResearch.4.L022060,
  title = {Observation of emergent ${\mathbb{Z}}_{2}$ gauge invariance in a superconducting circuit},
  author = {Wang, Zhan and Ge, Zi-Yong and Xiang, Zhongcheng and Song, Xiaohui and Huang, Rui-Zhen and Song, Pengtao and Guo, Xue-Yi and Su, Luhong and Xu, Kai and Zheng, Dongning and Fan, Heng},
  journal = {Phys. Rev. Res.},
  volume = {4},
  issue = {2},
  pages = {L022060},
  numpages = {6},
  year = {2022},
  month = {Jun},
  publisher = {American Physical Society},
  doi = {10.1103/PhysRevResearch.4.L022060},
  url = {https://link.aps.org/doi/10.1103/PhysRevResearch.4.L022060}
}

@article{meth_simulating_2025,
	title = {Simulating two-dimensional lattice gauge theories on a qudit quantum computer},
	volume = {21},
	issn = {1745-2481},
	url = {https://doi.org/10.1038/s41567-025-02797-w},
	doi = {10.1038/s41567-025-02797-w},
	number = {4},
	journal = {Nat. Phys.},
	author = {Meth, Michael and Zhang, Jinglei and Haase, Jan F. and Edmunds, Claire and Postler, Lukas and Jena, Andrew J. and Steiner, Alex and Dellantonio, Luca and Blatt, Rainer and Zoller, Peter and Monz, Thomas and Schindler, Philipp and Muschik, Christine and Ringbauer, Martin},
	month = apr,
	year = {2025},
	pages = {570--576},
}

@article{mueller_quantum_2025,
	title = {Quantum computing universal thermalization dynamics in a (2~+~1){D} {Lattice} {Gauge} {Theory}},
	volume = {16},
	issn = {2041-1723},
	url = {https://doi.org/10.1038/s41467-025-60177-7},
	doi = {10.1038/s41467-025-60177-7},
	number = {1},
	journal = {Nat. Commun.},
	author = {Mueller, Niklas and Wang, Tianyi and Katz, Or and Davoudi, Zohreh and Cetina, Marko},
	month = jul,
	year = {2025},
	pages = {5492},
}

@article{bauer2023quantum,
  title={Quantum simulation for high-energy physics},
  author={Bauer, Christian W and Davoudi, Zohreh and Balantekin, A Baha and Bhattacharya, Tanmoy and Carena, Marcela and De Jong, Wibe A and Draper, Patrick and El-Khadra, Aida and Gemelke, Nate and Hanada, Masanori and others},
  journal={PRX quantum},
  volume={4},
  number={2},
  pages={027001},
  year={2023},
  publisher={APS}
}

@article{tan_domain-wall_2021,
	title = {Domain-wall confinement and dynamics in a quantum simulator},
	volume = {17},
	issn = {1745-2481},
	url = {https://doi.org/10.1038/s41567-021-01194-3},
	doi = {10.1038/s41567-021-01194-3},
	number = {6},
	journal = {Nat. Phys.},
	author = {Tan, W. L. and Becker, P. and Liu, F. and Pagano, G. and Collins, K. S. and De, A. and Feng, L. and Kaplan, H. B. and Kyprianidis, A. and Lundgren, R. and Morong, W. and Whitsitt, S. and Gorshkov, A. V. and Monroe, C.},
	month = jun,
	year = {2021},
	pages = {742--747},
}

@article{kokail_self-verifying_2019,
	title = {Self-verifying variational quantum simulation of lattice models},
	volume = {569},
	issn = {1476-4687},
	url = {https://doi.org/10.1038/s41586-019-1177-4},
	doi = {10.1038/s41586-019-1177-4},
	number = {7756},
	journal = {Nature},
	author = {Kokail, C. and Maier, C. and van Bijnen, R. and Brydges, T. and Joshi, M. K. and Jurcevic, P. and Muschik, C. A. and Silvi, P. and Blatt, R. and Roos, C. F. and Zoller, P.},
	month = may,
	year = {2019},
	pages = {355--360},
}

@article{martinez_real-time_2016,
	title = {Real-time dynamics of lattice gauge theories with a few-qubit quantum computer},
	volume = {534},
	issn = {1476-4687},
	url = {https://doi.org/10.1038/nature18318},
	doi = {10.1038/nature18318},
	number = {7608},
	journal = {Nature},
	author = {Martinez, Esteban A. and Muschik, Christine A. and Schindler, Philipp and Nigg, Daniel and Erhard, Alexander and Heyl, Markus and Hauke, Philipp and Dalmonte, Marcello and Monz, Thomas and Zoller, Peter and Blatt, Rainer},
	month = jun,
	year = {2016},
	pages = {516--519},
}

@article{Klco_2022,
doi = {10.1088/1361-6633/ac58a4},
year = {2022},
month = {may},
publisher = {IOP Publishing},
volume = {85},
number = {6},
pages = {064301},
author = {Klco, Natalie and Roggero, Alessandro and Savage, Martin J},
title = {Standard model physics and the digital quantum revolution: thoughts about the interface},
journal = {Reports on Progress in Physics}
}

@article{Bali2005observation,
  title = {Observation of string breaking in QCD},
  author = {Bali, Gunnar S. and Neff, Hartmut and D\"ussel, Thomas and Lippert, Thomas and Schilling, Klaus},
  collaboration = {SESAM Collaboration},
  journal = {Phys. Rev. D},
  volume = {71},
  issue = {11},
  pages = {114513},
  numpages = {26},
  year = {2005},
  month = {Jun},
  publisher = {American Physical Society},
  doi = {10.1103/PhysRevD.71.114513}
}

@article{GLIOZZIA199976,
title = {The confining string and its breaking in QCD},
journal = {Nuclear Physics B},
volume = {556},
number = {1},
pages = {76-88},
year = {1999},
issn = {0550-3213},
doi = {https://doi.org/10.1016/S0550-3213(99)00330-2},
author = {F. Gliozzia and P. Provero}
}

@article{Philipsen1998string,
  title = {String Breaking in Non-Abelian Gauge Theories with Fundamental Matter Fields},
  author = {Philipsen, Owe and Wittig, Hartmut},
  journal = {Phys. Rev. Lett.},
  volume = {81},
  issue = {19},
  pages = {4056--4059},
  numpages = {0},
  year = {1998},
  month = {Nov},
  publisher = {American Physical Society},
  doi = {10.1103/PhysRevLett.81.4056}
}

@article{Chen2021Emergent,
  title = {Emergent symmetries and slow quantum dynamics in a Rydberg-atom chain with confinement},
  author = {Chen, I-Chi and Iadecola, Thomas},
  journal = {Phys. Rev. B},
  volume = {103},
  issue = {21},
  pages = {214304},
  numpages = {13},
  year = {2021},
  month = {Jun},
  publisher = {American Physical Society},
  doi = {10.1103/PhysRevB.103.214304}
}

@Article{Kormos2017,
author={Kormos, Marton
and Collura, Mario
and Tak{\'a}cs, Gabor
and Calabrese, Pasquale},
title={Real-time confinement following a quantum quench to a non-integrable model},
journal={Nature Physics},
year={2017},
month={Mar},
day={01},
volume={13},
number={3},
pages={246-249},
issn={1745-2481},
doi={10.1038/nphys3934},
}

@Article{Bloch2012,
author={Bloch, Immanuel
and Dalibard, Jean
and Nascimb{\`e}ne, Sylvain},
title={Quantum simulations with ultracold quantum gases},
journal={Nature Physics},
year={2012},
month={Apr},
day={01},
volume={8},
number={4},
pages={267-276},
issn={1745-2481},
doi={10.1038/nphys2259}
}

@article{Alkofer_2007,
doi = {10.1088/0954-3899/34/7/S02},
year = {2007},
month = {may},
publisher = {},
volume = {34},
number = {7},
pages = {S3},
author = {Alkofer, R and Greensite, J},
title = {Quark confinement: the hard problem of hadron physics},
journal = {Journal of Physics G: Nuclear and Particle Physics}
}

@article{verdel2020real,
  title = {Real-time dynamics of string breaking in quantum spin chains},
  author = {Verdel, Roberto and Liu, Fangli and Whitsitt, Seth and Gorshkov, Alexey V. and Heyl, Markus},
  journal = {Phys. Rev. B},
  volume = {102},
  issue = {1},
  pages = {014308},
  numpages = {15},
  year = {2020},
  month = {Jul},
  publisher = {American Physical Society},
  doi = {10.1103/PhysRevB.102.014308}
}

@article{funcke2022towards,
  title={Towards quantum simulations in particle physics and beyond on noisy intermediate-scale quantum devices},
  author={Funcke, Lena and Hartung, Tobias and Jansen, Karl and K{\"u}hn, Stefan and Schneider, Manuel and Stornati, Paolo and Wang, Xiaoyang},
  journal={Philosophical Transactions of the Royal Society A},
  volume={380},
  number={2216},
  pages={20210062},
  year={2022},
  publisher={The Royal Society}
}

@article{aidelsburger2022cold,
  title={Cold atoms meet lattice gauge theory},
  author={Aidelsburger, Monika and Barbiero, Luca and Bermudez, Alejandro and Chanda, Titas and Dauphin, Alexandre and Gonz{\'a}lez-Cuadra, Daniel and Grzybowski, Przemys{\l}aw R and Hands, Simon and Jendrzejewski, Fred and J{\"u}nemann, Johannes and others},
  journal={Philosophical Transactions of the Royal Society A: Mathematical, Physical and Engineering Sciences},
  volume={380},
  number={2216},
  year={2022},
  publisher={The Royal Society}
}

@Article{Banuls2020,
author={Ba{\~{n}}uls, Mari Carmen
and Blatt, Rainer
and Catani, Jacopo
and Celi, Alessio
and Cirac, Juan Ignacio
and Dalmonte, Marcello
and Fallani, Leonardo
and Jansen, Karl
and Lewenstein, Maciej
and Montangero, Simone
and Muschik, Christine A.
and Reznik, Benni
and Rico, Enrique
and Tagliacozzo, Luca
and Van Acoleyen, Karel
and Verstraete, Frank
and Wiese, Uwe-Jens
and Wingate, Matthew
and Zakrzewski, Jakub
and Zoller, Peter},
title={Simulating lattice gauge theories within quantum technologies},
journal={The European Physical Journal D},
year={2020},
month={Aug},
day={04},
volume={74},
number={8},
pages={165},
issn={1434-6079},
doi={10.1140/epjd/e2020-100571-8}
}

@article{zohar_quantum_2015,
	title = {Quantum simulations of lattice gauge theories using ultracold atoms in optical lattices},
	volume = {79},
	issn = {0034-4885},
	url = {https://doi.org/10.1088/0034-4885/79/1/014401},
	doi = {10.1088/0034-4885/79/1/014401},
	number = {1},
	urldate = {2021-07-08},
	journal = {Rep. Prog. Phys.},
	author = {Zohar, Erez and Cirac, J. Ignacio and Reznik, Benni},
	month = dec,
	year = {2015},
	pages = {014401},
}

@article{jordan_quantum_2012,
	title = {Quantum {Algorithms} for {Quantum} {Field} {Theories}},
	volume = {336},
	url = {https://www.science.org/doi/10.1126/science.1217069},
	doi = {10.1126/science.1217069},
	number = {6085},
	urldate = {2025-09-17},
	journal = {Science},
	author = {Jordan, Stephen P. and Lee, Keith S. M. and Preskill, John},
	month = jun,
	year = {2012},
	pages = {1130--1133},
}

@article{banerjee_atomic_2012,
	title = {Atomic {Quantum} {Simulation} of {Dynamical} {Gauge} {Fields} {Coupled} to {Fermionic} {Matter}: {From} {String} {Breaking} to {Evolution} after a {Quench}},
	volume = {109},
	shorttitle = {Atomic {Quantum} {Simulation} of {Dynamical} {Gauge} {Fields} {Coupled} to {Fermionic} {Matter}},
	url = {https://link.aps.org/doi/10.1103/PhysRevLett.109.175302},
	doi = {10.1103/PhysRevLett.109.175302},
	number = {17},
	urldate = {2025-10-07},
	journal = {Phys. Rev. Lett.},
	author = {Banerjee, D. and Dalmonte, M. and Müller, M. and Rico, E. and Stebler, P. and Wiese, U.-J. and Zoller, P.},
	month = oct,
	year = {2012},
	pages = {175302},
}

@article{tagliacozzo_optical_2013,
	title = {Optical {Abelian} lattice gauge theories},
	volume = {330},
	issn = {0003-4916},
	url = {https://www.sciencedirect.com/science/article/pii/S0003491612001819},
	doi = {10.1016/j.aop.2012.11.009},
	urldate = {2025-10-07},
	journal = {Ann. Phys. (N. Y.)},
	author = {Tagliacozzo, L. and Celi, A. and Zamora, A. and Lewenstein, M.},
	month = mar,
	year = {2013},
	pages = {160--191},
}

@article{zohar_simulating_2012,
	title = {Simulating {Compact} {Quantum} {Electrodynamics} with {Ultracold} {Atoms}: {Probing} {Confinement} and {Nonperturbative} {Effects}},
	volume = {109},
	shorttitle = {Simulating {Compact} {Quantum} {Electrodynamics} with {Ultracold} {Atoms}},
	url = {https://link.aps.org/doi/10.1103/PhysRevLett.109.125302},
	doi = {10.1103/PhysRevLett.109.125302},
	number = {12},
	urldate = {2025-10-07},
	journal = {Phys. Rev. Lett.},
	author = {Zohar, Erez and Cirac, J. Ignacio and Reznik, Benni},
	month = sep,
	year = {2012},
	pages = {125302},
}

@article{kogut_introduction_1979,
	title = {An introduction to lattice gauge theory and spin systems},
	volume = {51},
	copyright = {http://link.aps.org/licenses/aps-default-license},
	issn = {0034-6861},
	url = {https://link.aps.org/doi/10.1103/RevModPhys.51.659},
	doi = {10.1103/RevModPhys.51.659},
	number = {4},
	urldate = {2025-10-09},
	journal = {Rev. Mod. Phys.},
	author = {Kogut, John B.},
	month = oct,
	year = {1979},
	pages = {659--713},
}

@article{kogut_hamiltonian_1975,
	title = {Hamiltonian formulation of {Wilson}'s lattice gauge theories},
	volume = {11},
	copyright = {http://link.aps.org/licenses/aps-default-license},
	issn = {0556-2821},
	url = {https://link.aps.org/doi/10.1103/PhysRevD.11.395},
	doi = {10.1103/PhysRevD.11.395},
	number = {2},
	urldate = {2025-10-09},
	journal = {Phys. Rev. D},
	author = {Kogut, John and Susskind, Leonard},
	month = jan,
	year = {1975},
	pages = {395--408},
}

@article{yang_probing_2025,
	title = {Probing {Hilbert} space fragmentation with strongly interacting {Rydberg} atoms},
	volume = {111},
	issn = {2469-9950, 2469-9969},
	url = {https://link.aps.org/doi/10.1103/PhysRevB.111.144313},
	doi = {10.1103/PhysRevB.111.144313},
	number = {14},
	urldate = {2025-10-11},
	journal = {Phys. Rev. B},
	author = {Yang, Fan and Yarloo, Hadi and Zhang, Hua-Chen and M{\o}lmer, Klaus and Nielsen, Anne E. B.},
	month = apr,
	year = {2025},
	pages = {144313},
}

@article{surace_lattice_2020,
	title = {Lattice {Gauge} {Theories} and {String} {Dynamics} in {Rydberg} {Atom} {Quantum} {Simulators}},
	volume = {10},
	issn = {2160-3308},
	url = {https://link.aps.org/doi/10.1103/PhysRevX.10.021041},
	doi = {10.1103/PhysRevX.10.021041},
	number = {2},
	urldate = {2025-10-07},
	journal = {Phys. Rev. X},
	author = {Surace, Federica M. and Mazza, Paolo P. and Giudici, Giuliano and Lerose, Alessio and Gambassi, Andrea and Dalmonte, Marcello},
	month = may,
	year = {2020},
	pages = {021041},
}

@misc{darbha_probing_2025,
      title={Probing emergent prethermal dynamics and resonant melting on a programmable quantum simulator},
      author={Darbha, Siva and Khudorozhkov, Alexey and Lopes, Pedro L. S. and Liu, Fangli and Rrapaj, Ermal and Balewski, Jan and Hamdan, Majd and Dolgirev, Pavel E. and Schuckert, Alexander and Klymko, Katherine and Wang, Sheng-Tao and Lukin, Mikhail D. and Camps, Daan and Kornjača, Milan},
      eprint={2510.11706},
      archivePrefix={arXiv}
}

% \clearpage
\onecolumngrid
\vspace{30pt}

\begin{center}
\textbf{\large End Matter}
\end{center}    
\twocolumngrid

{\it Metastable confinement and prethermal dynamics.}---
The phenomenon of {metastable} confinement here is very similar to the prethermal dynamics. They are both constraints of Hilbert space ergodicity induced by a large energy scale. Here we have three different energy scales
\begin{equation}
    V_{1}\gg V_{2}\gg \Omega.
\end{equation}
The constraint induced by $V_{1}$ is from the total Hilbert space to the blockade Hilbert space. This is referred to as prethermal dynamics, because there is still limited ergodicity \cite{darbha_probing_2025}. $V_{2}$ introduces another stronger constraint from the blockade Hilbert space to a single state, which we referred to as confinement. They can be solved by the same theory \cite{berges_prethermalization_2004,abanin_rigorous_2017,mori_thermalization_2018}, which gives thermalization times $\mathcal{O}(e^{(V_{1}/V_{2})})$ and $\mathcal{O}(e^{(V_{2}/\Omega)})$, respectively. These two thermalization times can be similar, so the prethermal dynamics in the blockade Hilbert space may not exist, and the initial state can directly melt into the full Hilbert space. We only consider the blockade Hilbert space to calculate the thermal equilibrium, because the resonant melting can only induce ergodicity in the blockade subspace.

{\it Effective temperature.}---
The effective temperature $\beta$ is obtained by the following equations
\begin{equation}
    \mathrm{Tr}(\rho_{\mathrm{\beta}}H)=\langle Z_{2}|H|Z_{2}\rangle,
\end{equation}
\begin{equation}
    \rho_{\mathrm{\beta}}=\frac{e^{-\beta H}}{\mathrm{Tr}e^{-\beta H}},
\end{equation}
where $\rho_{\mathrm{\beta}}$ is the thermal state with temperature $\beta$ in the blockade Hilbert space. After determining the effective temperature, we can obtain the thermal expectation values of the nearest-neighbor correlation $\hat{O}_{ZZ}$ from the thermal state $\rho_{\mathrm{\beta}}$.

{\it Resonant subspace.}---
In the lower panel of Fig. \ref{fig:main}(a), we investigate the thermal equilibrium of both the blockade and resonant subspace. The first one is a subspace of all the states that satisfy the Rydberg blockade effect. The second one is a subspace of the states containing only one kind of excitation islands.

To be clear, we give an example of the second thermal equilibrium here. We consider the case $n=1$, which is also what we consider in Fig. \ref{fig:main}(a). The island is $``0100010"$. Consequently, for a 10-atom periodic array, the states in the resonant subspace are:
\begin{equation}
    \begin{split}
        \text{One-island states}:&1010101000\\
        &1010100010\\
        &1010001010\\
        &1000101010\\
        &0010101010\\
        \text{Two-island states}:&1010001000\\
        &1000101000\\
        &1000100010\\
        &0010100010\\
        &0010001010.
    \end{split}
\end{equation}
There are 5 one-island states and 5 two-island states. Larger number of islands is not supported. The total dimension of the resonant subspace is 10.

\begin{figure}[t]
    \centering
    \includegraphics[width=0.49\textwidth]{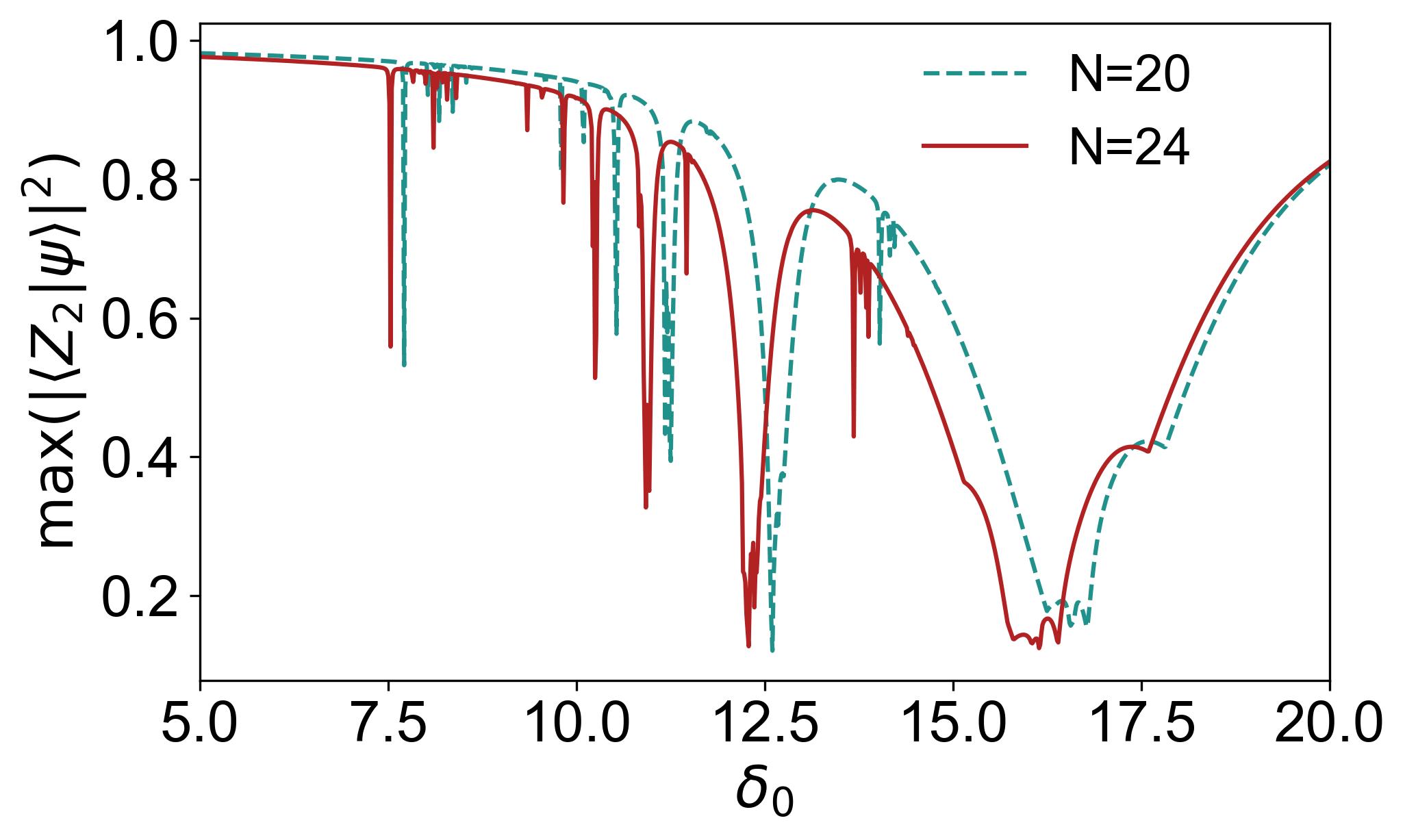}
    \caption{Maximal overlap between the eigenstates $|\psi_{i}\rangle$ and the initial state $|Z_{2}\rangle$ for two different atom numbers $N$.}
    \label{fig:fsize}
\end{figure}

{\it Finite-size effect.}---
The phenomena investigated here are not related to the system size. The finite-size effect originates from the circular periodic configuration of the Rydberg atoms. In this geometry, the distance between the $j$th atom and the $k$th atom is given by
\begin{equation}
    r_{jk}=2R\sin\left(\frac{|j-k|}{N}\pi\right),
\end{equation}
where $R$ is the radius of the circle array and $N$ is the number of atoms. Consequently, atom separation $a$ is
\begin{equation}
    a=2R\sin\left(\frac{1}{N}\pi\right),
\end{equation}
and then the atom distance can be rewritten as
\begin{equation}
    r_{jk}/a=\frac{\sin\left(|j-k|\pi/N\right)}{\sin\left(\pi/N\right)}.
\end{equation}
It gives the Rydberg interaction strengths as
\begin{equation}
    V_{jk}=\frac{C_{6}}{r_{jk}^{6}}=\Omega \left(\frac{R_{\mathrm{b}}}{a}\frac{\sin\left(\pi/N\right)}{\sin\left(|j-k|\pi/N\right)}\right)^{6}.
\end{equation}
They are functions of system size $N$ except for the nearest-neighbor interaction. Figure \ref{fig:fsize} shows the influence of system size on maximal overlap between eigenstates $|\psi_{i}\rangle$ and initial state $|Z_{2}\rangle$.  Although the resonances are shifted due to the finite-size effect, the main structure remains the same, and the difference will reduce as the system size becomes sufficiently large. There are more resonances because we improve the resolution of both lines. Many resonances are sharp because the dimension of the resonant Hilbert space is small. For simplicity, we do not show these sharp resonances in the main figures. They are automatically neglected when we reduce the line resolution.

{\it Towards experimental realization.}---
A challenge of our setup is that the required NNN interaction $V_{2}$ becomes too large to implement when the atoms are arranged in a regular geometry, such as a straight line or a circle. Fortunately, this constraint can be relaxed by placing the atoms in a zigzag configuration, thereby enhancing the NNN interaction. With an appropriately engineered structure, we can achieve the desired hierarchy of energy scales as $V_{1}\gg V_{2}\gg \Omega \gg V_{n>3}$.

Importantly, the underlying mechanism of metastable confinement and resonant string breaking can also be extended to a two‑dimensional U(1) lattice gauge theory by employing a mapping on a square \cite{celi_emerging_2020} or honeycomb lattice \cite{González-Cuadra2025}. In these cases, the ratios of the nearest-neighbor and the NNN interaction strengths are reduced, naturally allowing for the realization of large NNN interactions.

\end{document}